\documentclass[12pt,preprint]{aastex63}

\usepackage{amsmath}

\newcommand{\siiv}{Si {\sc iv}}
\newcommand{\fexxi}{Fe {\sc xxi}}

\shorttitle{Dynamic Processes in Limb Flares}
\shortauthors{Yu et al.}

\begin{document} 

\title{Imaging and Spectroscopic Observations of the Dynamic Processes in Limb Solar Flares}

\author{Ke Yu}           
\affiliation{School of Astronomy and Space Science, Nanjing University, Nanjing 210023, Peoples's Republic of China}
\affiliation{Key Laboratory for Modern Astronomy and Astrophysics (Nanjing University), Ministry of Education, Nanjing 210023, Peoples's Republic of China}

\author{Y. Li}
\affiliation{Key Laboratory of Dark Matter and Space Astronomy, Purple Mountain Observatory, Chinese Academy of Sciences, Nanjing 210033, Peoples's Republic of China}
\affiliation{School of Astronomy and Space Science, University of Science and Technology of China, Hefei 230026, Peoples's Republic of China}

\author{Jie Hong}           
\affiliation{School of Astronomy and Space Science, Nanjing University, Nanjing 210023, Peoples's Republic of China}
\affiliation{Key Laboratory for Modern Astronomy and Astrophysics (Nanjing University), Ministry of Education, Nanjing 210023, Peoples's Republic of China}

\author{De-Chao Song}
\affiliation{Key Laboratory of Dark Matter and Space Astronomy, Purple Mountain Observatory, Chinese Academy of Sciences, Nanjing 210033, Peoples's Republic of China}
\affiliation{School of Astronomy and Space Science, University of Science and Technology of China, Hefei 230026, Peoples's Republic of China}

\author{M. D. Ding}           
\affiliation{School of Astronomy and Space Science, Nanjing University, Nanjing 210023, Peoples's Republic of China}
\affiliation{Key Laboratory for Modern Astronomy and Astrophysics (Nanjing University), Ministry of Education, Nanjing 210023, Peoples's Republic of China}

\email{yingli@pmo.ac.cn, dmd@nju.edu.cn}

\begin{abstract}
We investigate various dynamic processes including magnetic reconnection, chromospheric evaporation, and coronal rain draining in two limb solar flares through imaging and spectroscopic observations from the Interface Region Imaging Spectrograph (IRIS) and the Atmospheric Imaging Assembly (AIA) on board the Solar Dynamics Observatory. In the early phase of the flares, a bright and dense loop-top structure with a cusp-like shape can be seen in multi-wavelength images, which is co-spatial with the hard X-ray 25--50 keV emission. In particular, intermittent magnetic reconnection downflows are detected in the time-space maps of AIA 304 \AA. The reconnection downflows are manifested as redshifts on one half of the loops and blueshifts on the other half in the IRIS \siiv\ 1393.76 \AA\ line due to a projection effect. The \siiv\ profiles exhibit complex features (say, multi-peak) with a relatively larger width at the loop-top region. During the impulsive phase, chromospheric evaporation is observed in both AIA images and the IRIS \fexxi\ 1354.08 \AA\ line. Upward motions can be seen from AIA 131 \AA\ images. The \fexxi\ line is significantly enhanced and shows a good Gaussian shape. In the gradual phase, warm rains are observed as downward moving plasmas in AIA 304 \AA\ images. Both the \siiv\ and \fexxi\ lines show a relatively symmetric shape with a larger width around the loop top. These results provide observational evidence for various dynamic processes involved in and are crucial to understand the energy release process of solar flares.
\end{abstract}

\keywords{Solar activity (1475); Solar flares (1496); Solar transition region (1532); Solar corona (1483); Solar ultraviolet emission (1533); Solar flare spectra (1982)}

\section{Introduction}

Solar flares are explosive energy release events in the solar atmosphere \citep[see a review by][]{flet11}. The standard flare model \citep{carm64, stur66, hira74, kopp76} suggests that they are triggered by magnetic reconnection occurring in the corona. The released energy first heats the local plasma and accelerates particles. Then the energy is transported along the newly reconnected loops to the lower atmosphere, where most of the energy is deposited. The chromosphere is heated and bright flare ribbons can be seen in H$\alpha$ and UV wavebands. The chromospheric heating can cause an excessive pressure, driving the plasma upward along the flare loops to the corona which is referred to as chromospheric evaporation \citep[e.g.,][]{fish85}. When the evaporation process has ceased, the plasma in the flare loops quickly cools down through radiative losses and thermal conduction, and thermal non-equilibrium leads to condensations in the corona \citep[e.g.,][]{ruan21}, which appear as dense blob-like plasma falling down along the loops, known as coronal rain \citep{anto12, oliv14, jing16, scul16, laca17}. 

In the past decades, there have been efforts to study the above different processes in solar flares through both imaging and spectroscopic observations. Firstly, the inflows and outflows of magnetic reconnection have widely been reported in imaging observations \citep[e.g.,][]{inne97, koyk03, linj05, mill10, nish10, wata12, taka12}. For example, \citet{liuw13} found high speed plasmoid ejections with velocities of 200--300 km s$^{-1}$, which represent bi-directional outflows, as observed by the Atmospheric Imaging Assembly \citep[AIA;][]{leme12} on board the Solar Dynamics Observatory \citep[SDO;][]{pesn12}. Combined with observations of the Reuven Ramaty High Energy Solar Spectroscopic Imager \citep[RHESSI;][]{linr02}, \citet{liuw13} also studied the particle acceleration and plasma heating in the outflows. In regard to spectroscopic observations before the era of  Interface Region Imaging Spectrograph \citep[IRIS;][]{depo14}, high blueshifts and redshifts in hot coronal lines have been reported at the loop-top region, interpreted as reconnection outflows \citep[e.g.,][]{inne03, wang07, hara11}. With the unprecedented resolution of IRIS, \citet{tian14} reported a large redshift in the Fe {\sc xxi} 1354.08 {\AA} line at the reconnection site of a C1.6 flare and interpreted it as the reconnection outflow. \citet{reev15} detected intermittent fast outflows from the \siiv~1393.76 {\AA} line as well as in AIA images. \citet{liyk17} presented the IRIS \siiv~1402.77 {\AA} line with a broadened width of 200 km s $^{-1}$ near the separator of an X-shaped flare, interpreted as reconnection bi-directional outflows. 

On the other hand, chromospheric evaporation and condensation are usually diagnosed with Doppler shifts and/or asymmetries of spectral lines. Generally speaking, the blueshifts of hot emission lines are a prominent signature of chromospheric evaporation. Before the IRIS era, there have been many observations of evaporation flows in flares \citep[e.g.,][]{bros03, bros04, harr05, mill06, grah11, youn13}, in which, however, the results suffered from the limited spatial resolution. Thanks to its high resolution, the evaporation flows can be much better resolved by IRIS. In particular, the Fe {\sc xxi} 1354.08 {\AA} line of IRIS has a formation temperature higher than 10 MK; thus the blueshifts of this line at flare ribbons are usually connected to chromospheric evaporation (see the review of \citealt{liyd19}). Meanwhile, redshifts of cool lines, e.g., Si {\sc iv}, C {\sc ii}, and Mg {\sc ii} lines, which are mainly formed in the chromosphere and transition region (TR), are good indications of chromospheric condensation. For example, \citet{tian15} tracked the complete evolution of evaporation flows in two flares and found an entirely blueshifted Fe {\sc xxi} 1354.08 {\AA} line and also red-wing enhanced cool lines like \siiv~1402.77 {\AA}. \citet{liyd19} studied gentle and explosive chromospheric evaporations in two flares. In the explosive evaporation, the cool Si {\sc iv}, C {\sc ii}, and Mg {\sc ii} lines show redshifts; while in the gentle one, these lines show blueshifts at the flare footpoints. 
 
Moreover, the redshifts in the cool lines are also widely detected in the gradual phase of flares. In this case, they often indicate downflows with a speed of tens to hundreds of km s$^{-1}$, many of which are supposed to be the result of warm rain due to cooling of the heated plasma \citep[e.g.,][]{bros03}. The coronal rain is mostly seen in TR and chromospheric lines, such as the \siiv, Ca {\sc ii}, and H$\alpha$ lines, and appears in post-flare loops \citep{bros03, bran16, scul16}. Reviews of coronal rains can be found in \citet{anto20} and \citet{lile21}.

Since the above phenomena are located at different heights of a flare loop, it would be very helpful to study them with limb observations \citep[e.g.,][]{leek20}. In this paper, we present spectroscopic and imaging observations for two limb flares observed by IRIS and AIA to exhibit different dynamic processes along flare loop structures. In Section \ref{sec-instr}, we describe the instruments and data reduction. Then we show detailed observational results in Section \ref{thi-res}. In Section \ref{fou-dis} we give the summary and discussions.

\section{Instruments and Data Reduction} 
\label{sec-instr}

The spectroscopic data presented here are from IRIS that provides high-resolution spectra in the far-ultraviolet (FUV, 1332--1358 {\AA}, 1389--1407 {\AA}) and near-ultraviolet (NUV, 2783--2834 {\AA}) wavelengths. The slit of IRIS has a width of 0.33$^{''}$ and can be used in a raster scan mode or a sit-and-stare mode. IRIS can also provide slit-jaw images (SJIs) at 1400 \AA, 1330 \AA, 2796 \AA, and 2832 {\AA} with a pixel scale of $0.166^{''}$. For the events under study, a large coarse 8-step raster was used and the slit scanned over a west limb region with an area of $14^{''} \times 119^{''}$. The exposure time at each step was $\sim$8 s, and each raster took about 74 s to complete. The SJIs were recorded at 1400 {\AA} and 1330 {\AA} with a field of view of $119^{''} \times 119^{''}$ and a cadence of 37 s. The level 2 data are used, which have been calibrated including subtraction of dark current and correction for flat field, geometry, and wavelength. 

In this work, we mainly study the \siiv~line at 1393.76 {\AA} with a formation temperature of $\sim$10$^{4.9}$ K and also the \fexxi\ line at 1354.08 {\AA} with a formation temperature of $\sim$10$^{7.0}$ K. Considering that these two lines show some complex profiles (say, asymmetric or multi-peak ones) during the limb flares, here we apply a moment analysis to obtain the spectral parameters, in which the zeroth, first, and second order moments correspond to the integrated or total line intensity ($I_t$), line center ($\lambda_c$), and equivalent line width ($w$), respectively. The three parameters can be expressed as follows (also see \citealt{yuke20}):

\begin{equation}
I_t=\int\,I(\lambda)\,d\lambda,
\end{equation}
\begin{equation}
\lambda_c=\left[\int\lambda\,I(\lambda)\,d\lambda\right]/I_t,
\end{equation}
 and
\begin{equation}
w^2=\left[\int(\lambda-\lambda_c)^2\,I(\lambda)\,d\lambda\right]/I_t,
\end{equation}
where $I(\lambda)$ is the line intensity at wavelength $\lambda$. The line shift or the Doppler velocity ($v$) can be derived by $v=c(\lambda_c-\lambda_r)/\lambda_r$, where $c$ is the light speed and $\lambda_r$ represents the reference wavelength of the line. For $\lambda_r$, here we adopt the theoretical or calibrated wavelengths of 1393.755 {\AA} and 1354.080 {\AA} for the \siiv\ and \fexxi\ lines, respectively. On the one hand, we understand that the IRIS data have been made the wavelength calibration by using some photospheric lines, which has an accuracy of $\sim$1 km s$^{-1}$ \citep{depo14}. On the other hand, we also check these reference wavelengths by averaging the line profiles over some regions for our events. For \siiv, we select a relatively quiet region on the solar disk (during 22:50--23:30 UT) and obtain an average line center of 1393.744 \AA. Regarding \fexxi, we average the profiles over the flaring region but in the late decay phase (02:00--02:15 UT) and obtain a line center of 1354.094 \AA. These two wavelength values are very similar to the above reference wavelengths, only with a difference of 2 or 3 km s$^{-1}$. Note that all of the line profiles analyzed in this work do not suffer a saturation effect. In contrast, some profiles (mainly outside the flaring loops above the limb) exhibit a relatively weak intensity. In order to improve the reliability, we then set an intensity threshold (three wavelength pixels in the line profile having intensities greater than 20 DNs and 15 DNs for \siiv\ and \fexxi, respectively) when making the moment analysis to the observed line profiles.

We also use the imaging data from the SDO/AIA. AIA provides full-disk images in one white-light (4500 {\AA}), two UV (1700 {\AA} and 1600 {\AA}) and seven EUV (335 \AA, 304 {\AA}, 211 {\AA}, 193 {\AA}, 171 {\AA}, 131 {\AA}, and 94 {\AA}) channels. The images have a pixel scale of 0.6$^{''}$ and a cadence of 12 s in EUV and 24 s in UV. Note that the AIA 304 \AA\ images are sensitive at $\sim$10$^{4.7}$ K, which show similar features, e.g., cool loops or flare ribbons, with the IRIS SJIs at 1400 {\AA} and also 1330 {\AA}. As for the AIA 131 {\AA} and 94 {\AA} images, we usually see some high-temperature structures like flare loops or fuzzy plasmoids. 

Some data from the RHESSI and the Geostationary Operational Environmental Satellites (GOES) related to the two flares are used in this study as well. RHESSI can provide X-ray and $\gamma$-ray spectra in the range from 3 keV up to 17 MeV with an energy resolution of 1--5 keV. The temporal resolution can be better than 2 s and the spatial resolution can be as fine as $2.3^{''}$. For the two flare events, we mainly show RHESSI 12--25 keV and 25--50 keV light curves. Some reconstructed images at 25--50 keV are also exhibited. For the GOES data, we present the 1--8 {\AA} light curves to show the classes and the global evolution of the flares.
 
\section{Analysis and Results}
\label{thi-res}

\subsection{Overview of the Limb Flare Events}

\begin{figure*}[ht!]
\epsscale{1.2}
\plotone{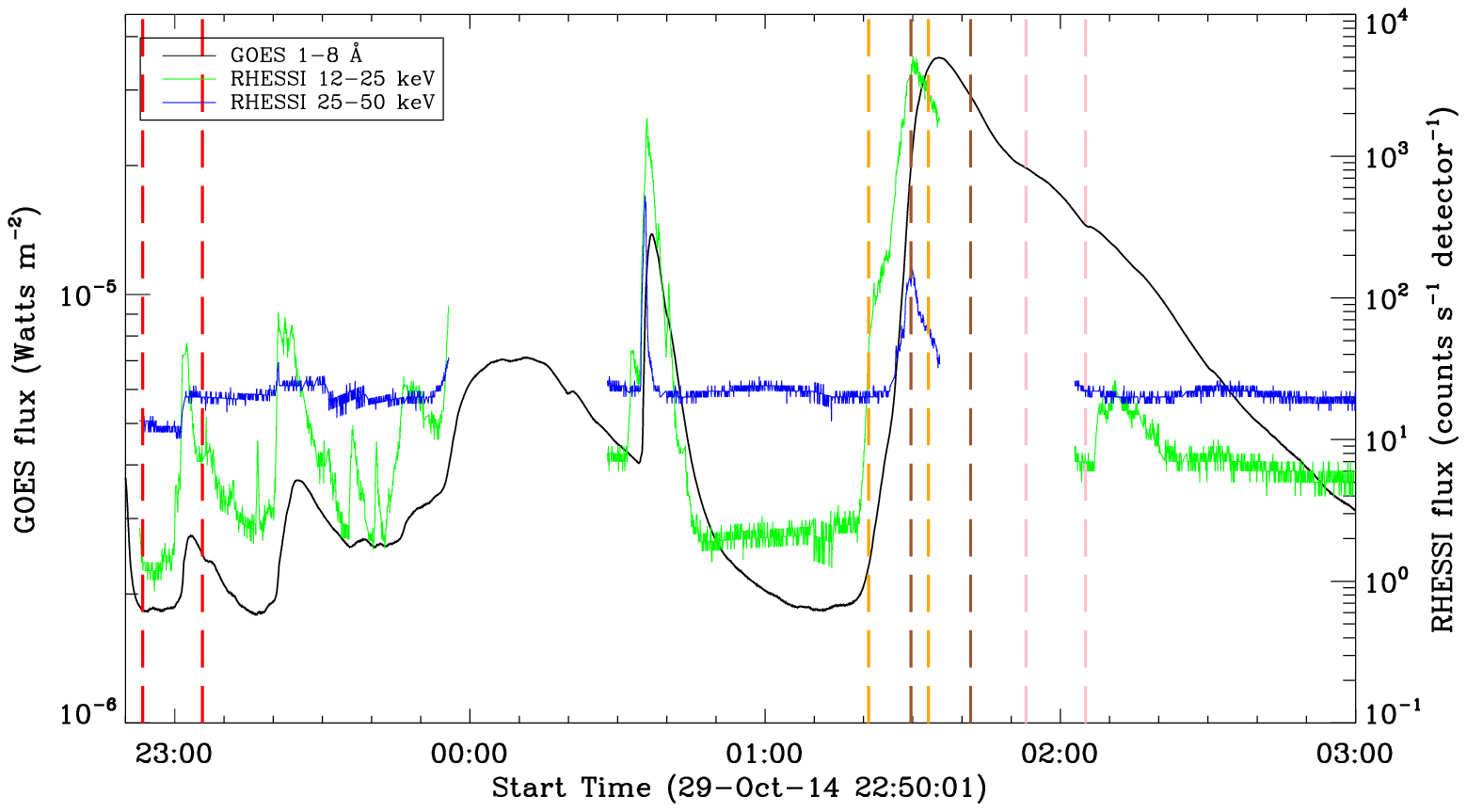}
\caption{Light curves of the GOES SXR and RHESSI HXR emissions. Four pairs of vertical dashed lines indicate the time periods we mainly focus on, where red, yellow, brown and pink correspond to the ones shown in Figures \ref{fig-2}(f) and \ref{fig-4}, \ref{fig-3}(f) and \ref{fig-5}, \ref{fig-8}(f) and \ref{fig-9}, and \ref{fig-11}(f) and \ref{fig-12}, respectively. Note that there are some gaps where the RHESSI data are unavailable.\label{fig-1}}
\end{figure*}

The flare events under study occurred in active region NOAA 12192, located at the west limb, on 2014 October 29 and 30. IRIS started to observe this limb region at 22:49 UT on 2014 October 29 and ended at 02:55 UT on 2014 October 30. During this period, a few C-class flares as well as two M-class flares took place, which can be seen from the GOES 1--8 {\AA} soft X-ray light curve in Figure \ref{fig-1}. We focus on two of them, a C2.7 flare around 23:00 UT on October 29 (see the two red vertical dashed lines) and an M3.5 flare around 01:30 UT on October 30 (marked by the other vertical dashed lines). The flares show up loop configurations above the solar limb, as clearly seen in SJIs at 1400 {\AA} and 1330 {\AA} as well as AIA 304 {\AA} images (see Figures \ref{fig-2} and \ref{fig-3}). In particular, the loop structures at different heights are crossed over by the IRIS slit with different scan steps, which provides a rare opportunity to study the dynamic processes in flare loops from a side view. From Figure \ref{fig-1} we can also see that the C-class flare has hardly response at RHESSI 25--50 keV while the M-class one shows an obvious emission at this energy band. In the following, we study various dynamic processes including magnetic reconnection, chromospheric evaporation, and coronal rain (or coronal condensation) in these two limb flares through imaging and spectroscopic observations. 

\subsection{Downflows from Magnetic Reconnection}
\label{thi-res1}

\begin{figure*}[ht!]
\epsscale{1.2}
\plotone{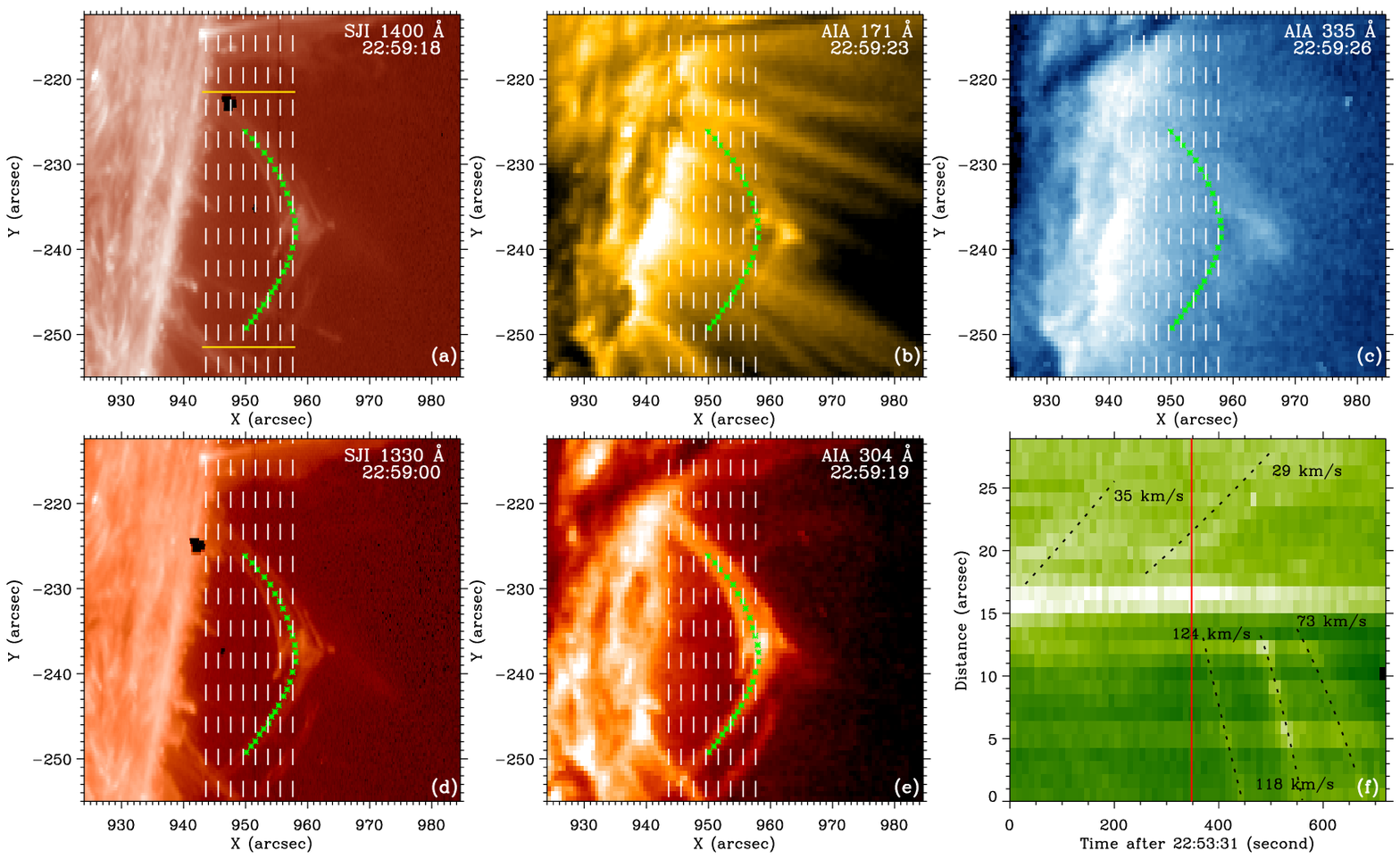}
\caption{
(a)--(e): IRIS SJI 1400 {\AA}, AIA 171 {\AA}, AIA 335 {\AA}, SJI 1330 {\AA} and AIA 304 {\AA} images taken at 22:59 UT on 2014 October 29 for the C2.7 flare. The white vertical dashed lines in these panels mark the eight scan steps of the IRIS slit. The two yellow horizontal lines in panel (a) denote the region as shown in Figure \ref{fig-4}. (f) Time-space map at AIA 304 {\AA} for the slice along a loop structure indicated by the green curve in panels (a)--(e). The red vertical line marks the time at 22:59 UT, corresponding to the one in IRIS and AIA images. Some measured velocities in the plane of the sky are given on the map.}
\label{fig-2}
\end{figure*}

Figures \ref{fig-2}(a)--(e) show multi-wavelength SJIs and AIA images at the onset time of the C2.7 flare ($\sim$22:59 UT, also denoted by the red vertical line in Figure \ref{fig-2}(f)). We can clearly see some loop structures (indicated by the green curve in the panels) especially in the low-temperature channels of 304 \AA, 1330 \AA, and 1400 \AA. It is also interesting that there are some bright and dense plasma structures near the loop-top region, which appear as a cusp-like shape and are seen more prominently in the relatively high-temperature channels of 171 \AA\ and 335 \AA. We track the temporal evolution of AIA 304 \AA\ emission along the loop slice (i.e., the green curve) and display the time-space diagram in Figure \ref{fig-2}(f). It is seen that some downward plasma motions from the loop-top can be detected around the flare onset (from 22:53--23:05 UT), whose speeds are from a few tens of to more than one hundred km s$^{-1}$ in the plane of the sky. Note that the downflow speeds become larger when the flare enters its rise phase. All these may suggest that magnetic reconnection takes place and produces the C-class flare.

\begin{figure*}[ht!]
\epsscale{1.2}
\plotone{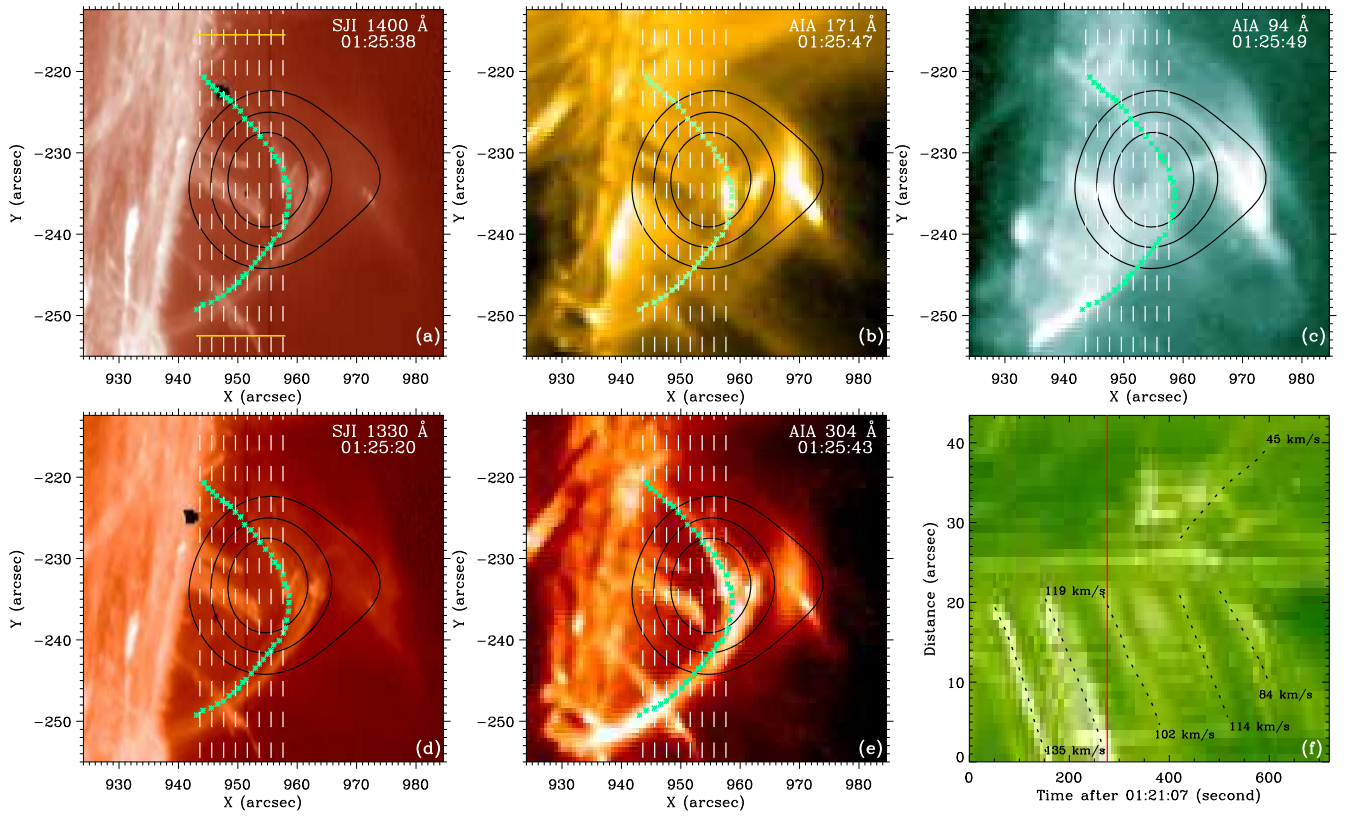}
\caption{
(a)--(e): IRIS SJI 1400 {\AA}, AIA 171 {\AA}, AIA 94 {\AA}, SJI 1330 {\AA}, and AIA 304 {\AA} images taken at 01:25 UT on 2014 October 30 for the M3.5 flare. The white vertical dashed lines in these panels mark the eight scan steps of the IRIS slit and the black contours denote the RHESSI 25--50 keV source reconstructed from 01:25:40 UT to 01:29:04 UT. The two yellow horizontal lines in panel (a) indicate the region as shown in Figure \ref{fig-5}. (f) Time-space map at AIA 304 {\AA} for the slice along a loop structure indicated by the green curve in panels (a)--(e). The red vertical line marks the time at 01:25 UT, corresponding to the one in IRIS and AIA images. Some measured velocities in the plane of the sky are given on the map.}
\label{fig-3}
\end{figure*}

Similarly, multi-wavelength SJIs, AIA images, and the time-space map of AIA 304 \AA\ for the M3.5 flare are shown in Figure \ref{fig-3}. We can clearly see bright loop structures in both low- and high-temperature channels during the rise phase of the flare (from 01:21--01:33 UT). Bright and dense structure at the loop top (and even above the loop top) can also be seen in these images. In addition, the RHESSI 25--50 keV contours are co-spatial well with the loop-top structures, both of which show a cusp-like shape. All of these observational features are consistent with the magnetic reconnection scenario as illustrated in the standard solar flare model. More interestingly, intermittent downflows with speeds of more than one hundred km s$^{-1}$ can be seen along the loop structure (marked by the green curve) from the time-space map of AIA 304 \AA, indicative of an intermittent magnetic reconnection. Note that due to the projection effect, more prominent reconnection downflows are detected in the southern halves of the loops.

\begin{figure*}[ht!]
\epsscale{1.2}
\plotone{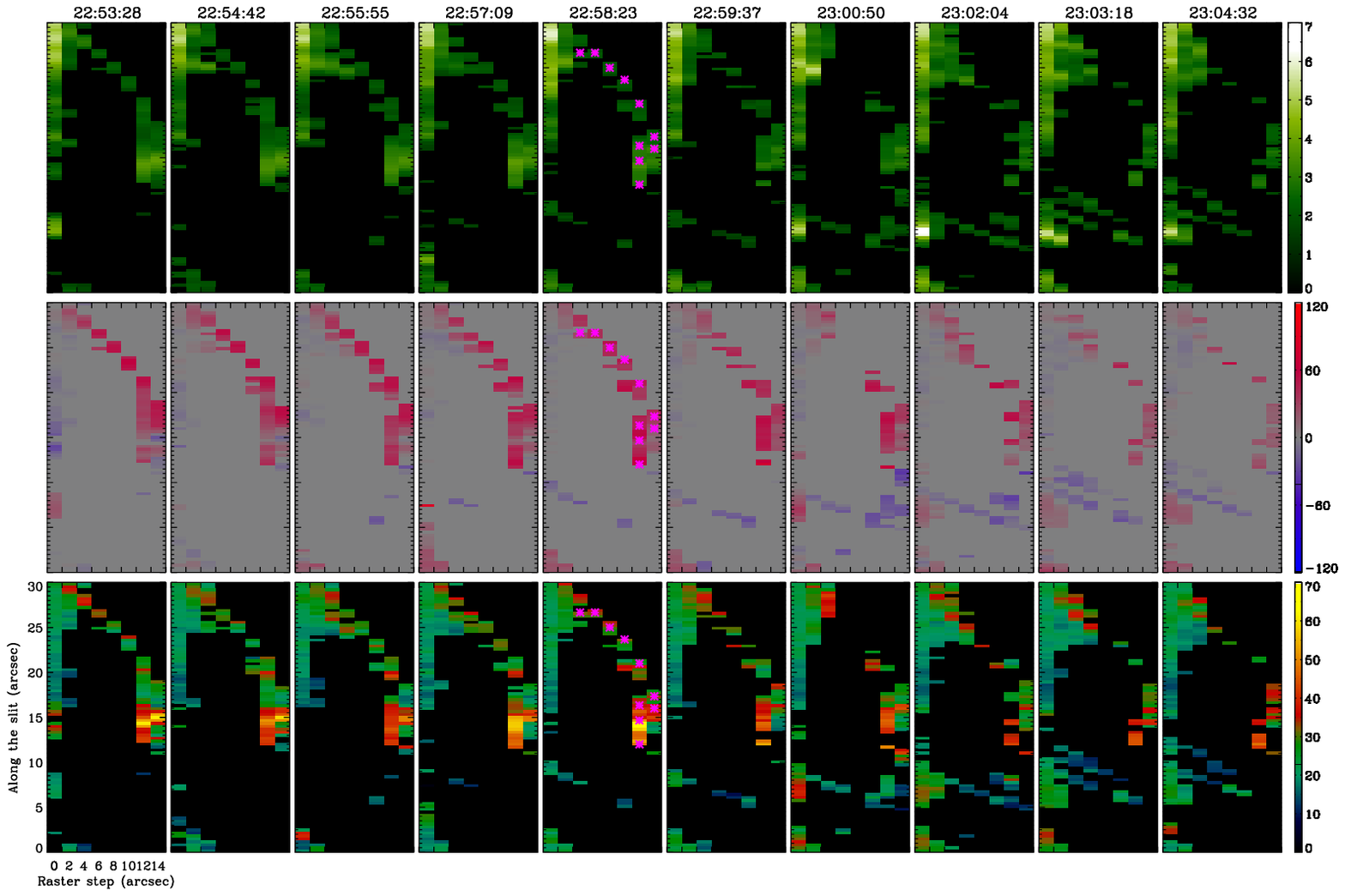}
\caption{
The maps of integrated intensity (top row, in units of log (DNs)), Doppler velocity (middle row, in units of km s$^{-1}$) and line width (bottom row, in units of km s$^{-1}$) of the \siiv~1393.76 {\AA} line obtained from moment analysis from 22:53 UT to 23:05 UT on 2014 October 29 with an interval of 74 s, which corresponds to the time period for the time-space map in Figure \ref{fig-2}(f). The starting times for each raster scan is given at the top of each column. The ten magenta asterisk symbols in the fifth column (corresponding to $\sim$22:59 UT) denote the locations along a half loop structure where the \siiv\ line profiles are exhibited in Figure \ref{fig-6}.}
\label{fig-4}
\end{figure*}

\begin{figure*}[ht!]
\epsscale{1.2}
\plotone{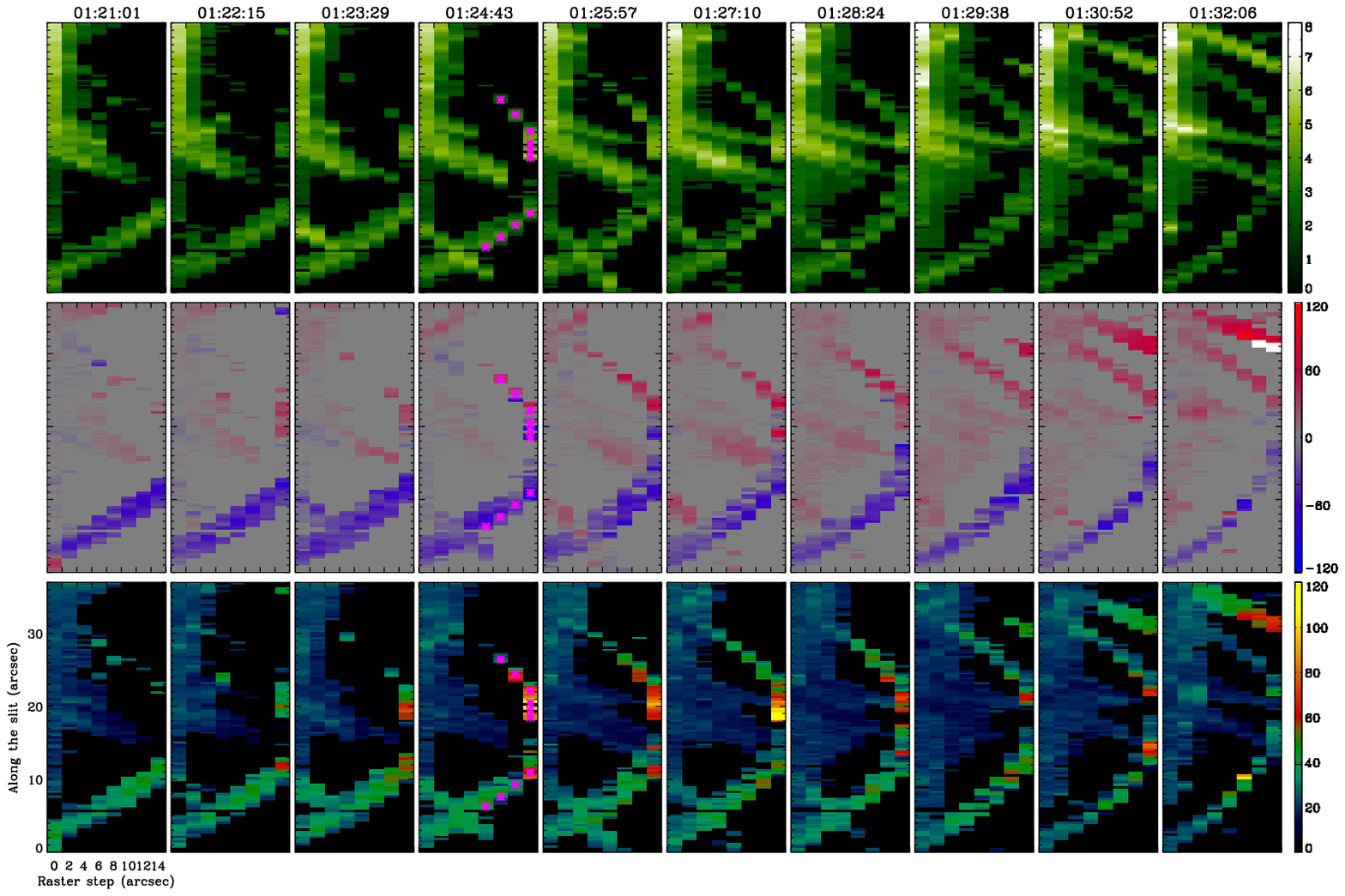}
\caption{
The maps of integrated intensity (top row, in units of log (DNs)), Doppler velocity (middle row, in units of km s$^{-1}$) and line width (bottom row, in units of km s$^{-1}$) of \siiv~1393.76 {\AA} line obtained from moment analysis from 01:21 UT to 01:33 UT on 2014 October 30 with an interval of 74 s, which corresponds to the time period for the time-space map in Figure \ref{fig-3}(f). The starting time for each raster scan is given at the top of each column. The ten magenta asterisk symbols in the fourth column (corresponding to $\sim$01:25 UT) denote the locations along a loop structure where the \siiv\ line profiles are exhibited in Figure \ref{fig-7}.}
\label{fig-5}
\end{figure*}

Figures \ref{fig-4} and \ref{fig-5} show the integrated intensity, Doppler velocity, and line width maps of the IRIS \siiv\ 1393.76 {\AA} line related to the two flares for ten raster scans corresponding to the time ranges in Figures \ref{fig-2}(f) and \ref{fig-3}(f), respectively. One can see the newly reconnected loop structures on these maps as well. Note that for the C-class flare shown in Figure \ref{fig-4}, its southern halves of the loops are fairly week where many pixels have intensity values below a certain threshold, while the loop-top region has mostly a fairly strong intensity. From the Doppler velocity maps, we can see that the northern halves of the loops in both flares show redshifts while the southern halves exhibit blueshifts in general, especially in the M-class flare as shown in Figure \ref{fig-5}. This is obviously due to the projection effect. Therefore we can suppose that the southern halves of the loops are closer to us in the line of sight. In general, the loop-top region has a larger redshift or blueshift velocity (from tens to hundreds of km s$^{-1}$) than the loop legs (about a few tens of km s$^{-1}$), which may imply that the reconnection downflows are decelerated when traveling downward along the loops. From the line width maps, it is seen that the loop-top region shows a significantly greater line width than the loop legs, which suggests that the loop-top region is likely more turbulent. It should be mentioned that the line widths in the M-class flare (up to $\sim$120 km s$^{-1}$) are much larger than the ones in the C-class flare (only up to $\sim$70 km s$^{-1}$). Moreover, the newly reconnected loops in both flares become higher and higher as flares proceed; thus the loop-top region cannot be well captured by the IRIS slit at a later time.

\begin{figure*}[ht!]
\epsscale{1.2}
\plotone{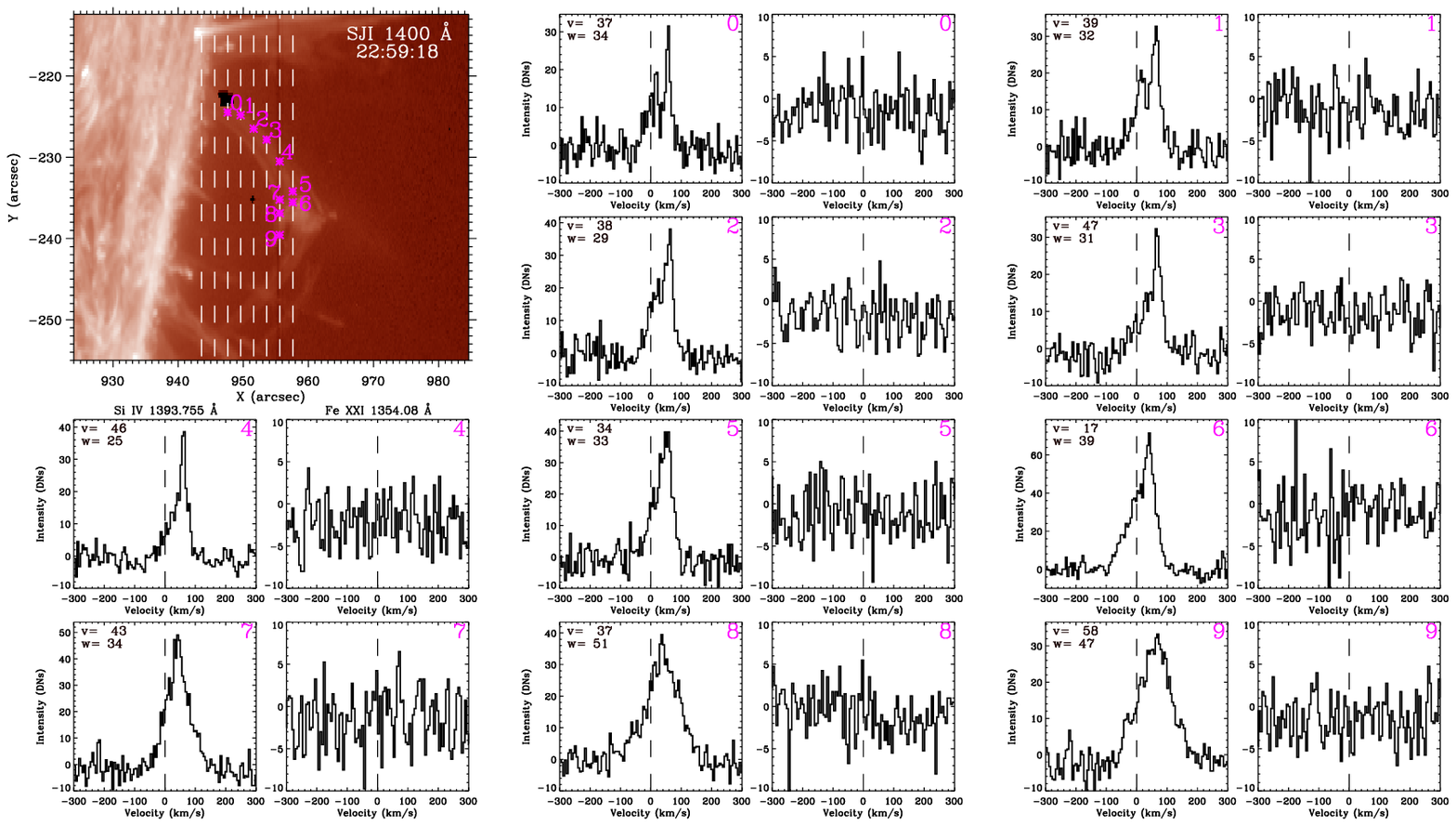}
\caption{
\siiv~1393.76 {\AA} and Fe {\sc xxi} 1354.08 {\AA} line profiles for ten different locations along a half loop structure as shown in the IRIS SJI 1400 {\AA} image at $\sim$22:59 UT on 2014 October 29. These locations correspond to the ones marked in the fifth column of Figure \ref{fig-4} and are labeled as 0--9. The vertical dashed line in the panels of line profiles indicates the reference wavelength of the \siiv\ or Fe {\sc xxi} line. The Doppler velocity and line width (in units of km s$^{-1}$) of the \siiv\ line derived from the moment analysis are presented. Note that the Fe {\sc xxi} line has no obvious response at these locations.}
\label{fig-6}
\end{figure*}

\begin{figure*}[ht!]
\epsscale{1.2}
\plotone{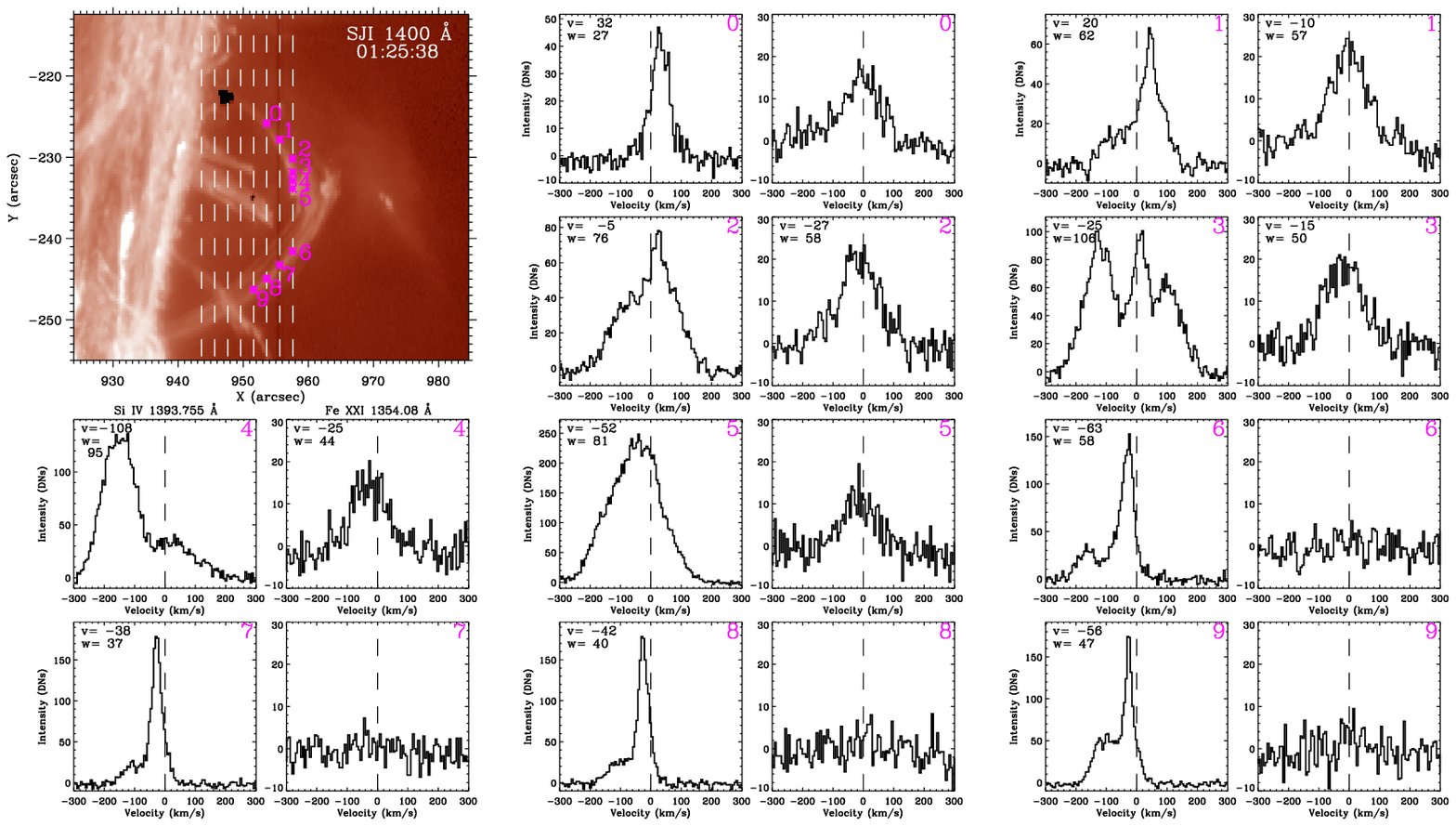}
\caption{
\siiv~1393.76 {\AA} and Fe {\sc xxi} 1354.08 {\AA} line profiles for ten different locations along a loop structure as shown in the IRIS SJI 1400 {\AA} image at $\sim$01:25 UT on 2014 October 30. These locations correspond to the ones marked in the fourth column of Figure \ref{fig-5} and are labeled as 0--9. The vertical dashed line in the panels of line profiles indicates the reference wavelength of the \siiv\ or Fe {\sc xxi} line. The Doppler velocity and line width (in units of km s$^{-1}$) derived from the moment analysis are presented. Note that the Fe {\sc xxi} line shows an enhancement at some locations.}
\label{fig-7}
\end{figure*}

We plot the line profiles of \siiv\ 1393.76 {\AA} and \fexxi\ 1354.08 {\AA} along the newly reconnected loop structure for the two flares in Figures \ref{fig-6} and \ref{fig-7}, which were obtained from one of the ten raster scans (see the magenta asterisk symbols in Figures \ref{fig-4} and \ref{fig-5}, corresponding to the time instant for Figures \ref{fig-2}(a)--(e) and that for Figures \ref{fig-3}(a)--(e), respectively). For the C-class flare in Figure \ref{fig-6}, we can see that the \siiv\ profiles mainly show multiple redshifted peaks or a red asymmetry with a relatively narrower line width ($<$40 km s$^{-1}$) on its northern loop legs (say, locations 0--6), while the profiles from the loop-top region (locations 7--9) look more symmetric but with a larger width ($\sim$50 km s$^{-1}$). The multi-peaks of the \siiv\ profile from the loop legs could originate from multiple loops overlapping along the line of sight and containing flows with different velocities. Regarding the relatively symmetric and wide \siiv\ profiles around the loop-top region, they are more likely contributed by turbulent flows associated with magnetic reconnection. Note that the hot \fexxi\ line has no obvious emission in the C-class flare, indicating that the plasma may not be heated to 10 MK. Unlike the C-class flare, the M-class flare has plasma being heated to 10 MK since the \fexxi\ line shows an emission mainly on its northern loop legs, i.e., locations 0--5, as seen in Figure \ref{fig-7}. The \siiv\ line also exhibits some interesting features during the M-class flare. On the loop legs (say, locations 0--2 and 6--9), the \siiv\ profiles show a slightly redshifted or blueshifted component with a relatively narrower width ($<$80 km s$^{-1}$), while the profiles near the loop-top region (i.e., locations 3--5) display multiple peaks, some of which have a strong velocity ($>$100 km s$^{-1}$) as well as a large width ($\sim$100 km s$^{-1}$).

\subsection{Upflows from Chromospheric Evaporation}
\label{thi-res2}

\begin{figure*}[ht!]
\epsscale{1.2}
\plotone{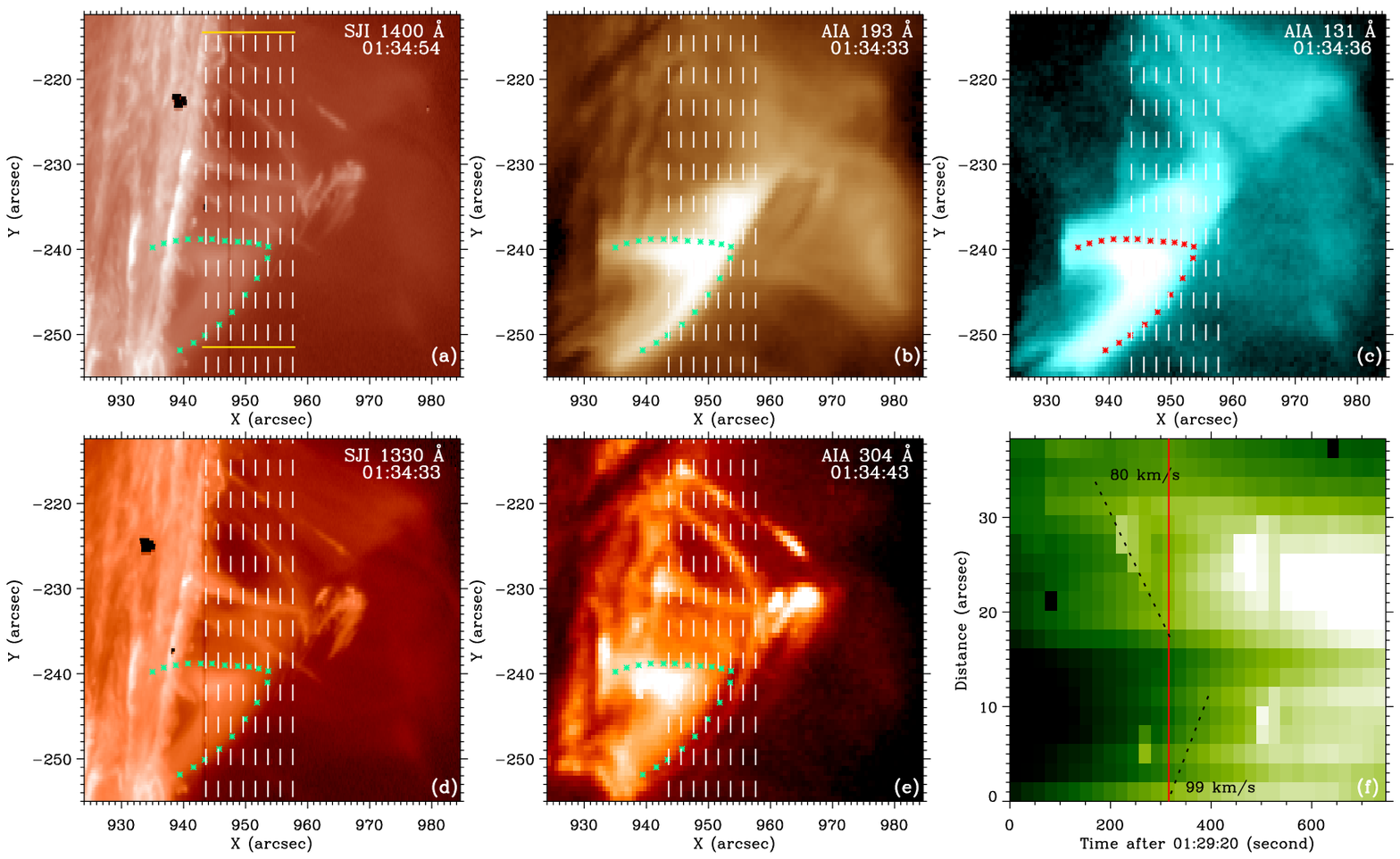}
\caption{
(a)--(e): IRIS SJI 1400 {\AA}, AIA 193 {\AA}, AIA 131 {\AA}, SJI 1330 {\AA}, and AIA 304 {\AA} images taken at 01:34 UT on 2014 October 30 for the M3.5 flare. The white vertical dashed lines in these panels mark the eight scan steps of the IRIS slit. The two yellow horizontal lines in panel (a) denote the region as shown in Figure \ref{fig-9}. (f) Time-space map at AIA 131 {\AA} for the slice along a loop structure indicated by the green or red curve in panels (a)--(e). The red vertical line marks the time at 01:34 UT, corresponding to the one in IRIS and AIA images. Some measured velocities in the plane of the sky are given on the map.}
\label{fig-8}
\end{figure*}

Figures \ref{fig-8}(a)--(e) show multi-wavelength SJIs and AIA images around the peak time of the M3.5 flare ($\sim$01:34 UT, also marked by the red vertical line in Figure \ref{fig-8}(f)). We can see prominent hot flare loops or post-reconnection loops (marked by the green or red curve) in AIA 193 \AA\ and 131 \AA\ channels that are sensitive to high temperatures of $\sim$10$^{7.3}$ K and 10$^{7.0}$ K, respectively. Note that some cooler and longer newly reconnected loops plus a bright emission source above the loop top can still be seen in the low-temperature channels of 304 \AA, 1330 \AA, and 1400 \AA\ at this time. The hot flare loops are mainly filled with heated plasmas that have been evaporated into the corona from the chromosphere. We plot the time-space diagram of AIA 131 \AA\ along the flare loop structure (i.e., along the green curve in Figures \ref{fig-8}(e), excluding the saturation area as possible as we can) for the main flare phase (from 01:29--01:41 UT) in Figure \ref{fig-8}(f). It is seen that hot plasmas continually move upward to the loop top from the footpoints with speeds of a few tens of km s$^{-1}$ in the plane of the sky. This provides an imaging evidence for chromospheric evaporation.

\begin{figure*}[ht!]
\epsscale{1.2}
\plotone{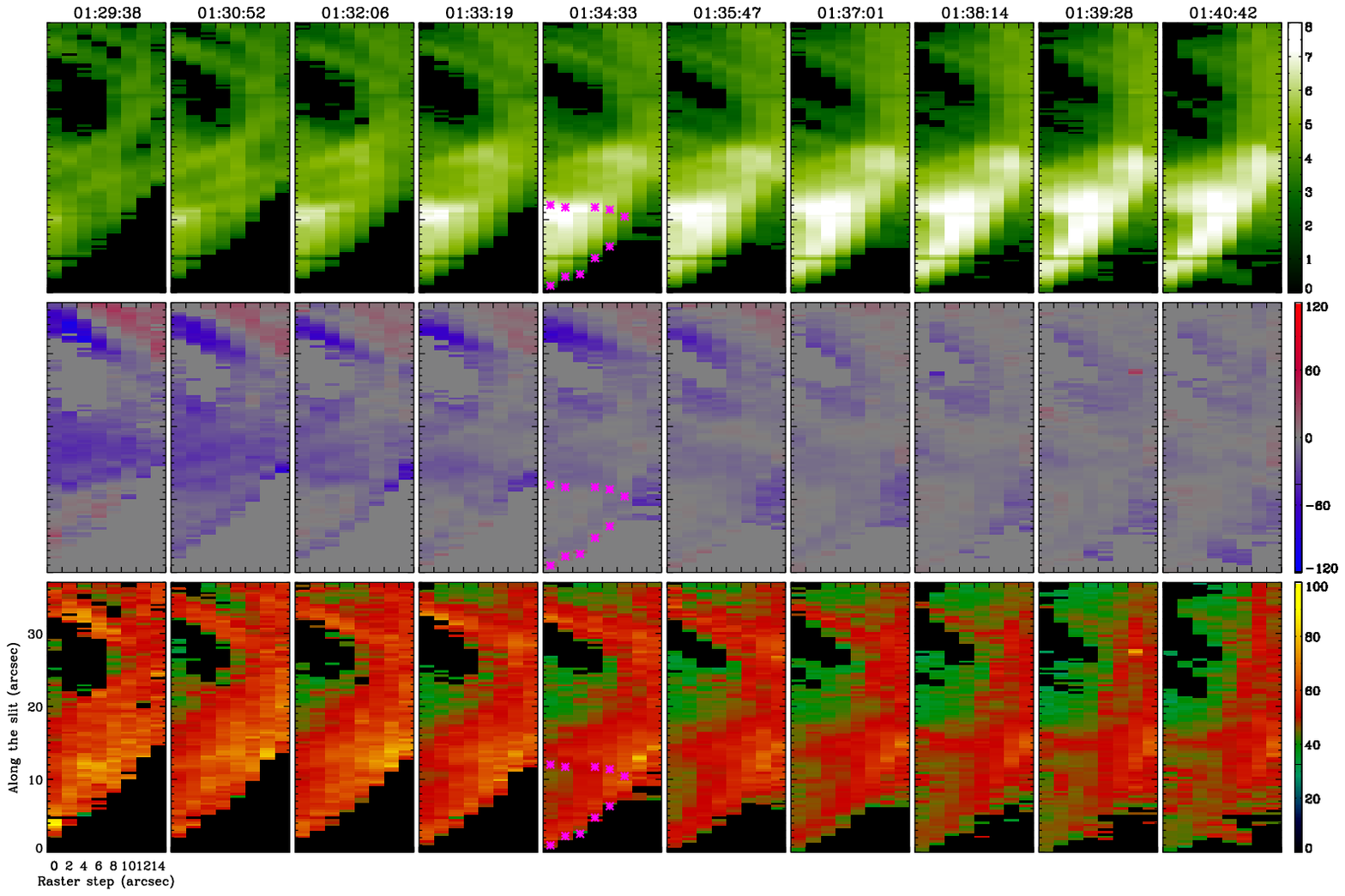}
\caption{
The maps of integrated intensity (top row, in units of log (DNs)), Doppler velocity (middle row, in units of km s$^{-1}$) and line width (bottom row, in units of km s$^{-1}$) of the Fe {\sc xxi} 1354.08 {\AA} line obtained from moment analysis from 01:29 UT to 01:41 UT on 2014 October 30 with an interval of 74 s, which corresponds to the time period for the time-space map in Figure \ref{fig-8}(f). The starting time for each raster scan is given at the top of each column. The ten magenta asterisk symbols in the fifth column (corresponding to $\sim$01:34 UT) denote the locations along a loop structure where the Fe {\sc xxi} line profiles are exhibited in Figure \ref{fig-10}.}
\label{fig-9}
\end{figure*}

We show the integrated intensity, Doppler velocity, and line width maps of the IRIS \fexxi\ 1354.08 {\AA} line for ten raster scans (corresponding to the time period in Figure \ref{fig-8}(f)) of the M-class flare in Figure \ref{fig-9}. One can see prominent flare loop structures (indicated by the magenta asterisk symbols) especially on the intensity and width maps. Compared with the newly reconnected loops as described above, the post-flare loops have much stronger emissions in the hot \fexxi\ line. The \fexxi\ line also exhibits some redshifts or blueshifts but with relatively small velocities ($<$60 km s$^{-1}$) at the post-flare loops. Note that the IRIS slit does not cross over the flare loop footpoints that usually exhibit more evident evaporation flows. In addition, considering the inclination of the flare loops, i.e., mostly in the plane of the sky, the evaporation flows along the flare loops could not be well detected in spectral lines from the line of sight. From the line width maps, it is seen that the loop-top region also has a larger width ($\sim$100 km s$^{-1}$) than the loop legs (a few tens of km s$^{-1}$), which implies that the plasma flows are more turbulent there.

\begin{figure*}[ht!]
\epsscale{1.2}
\plotone{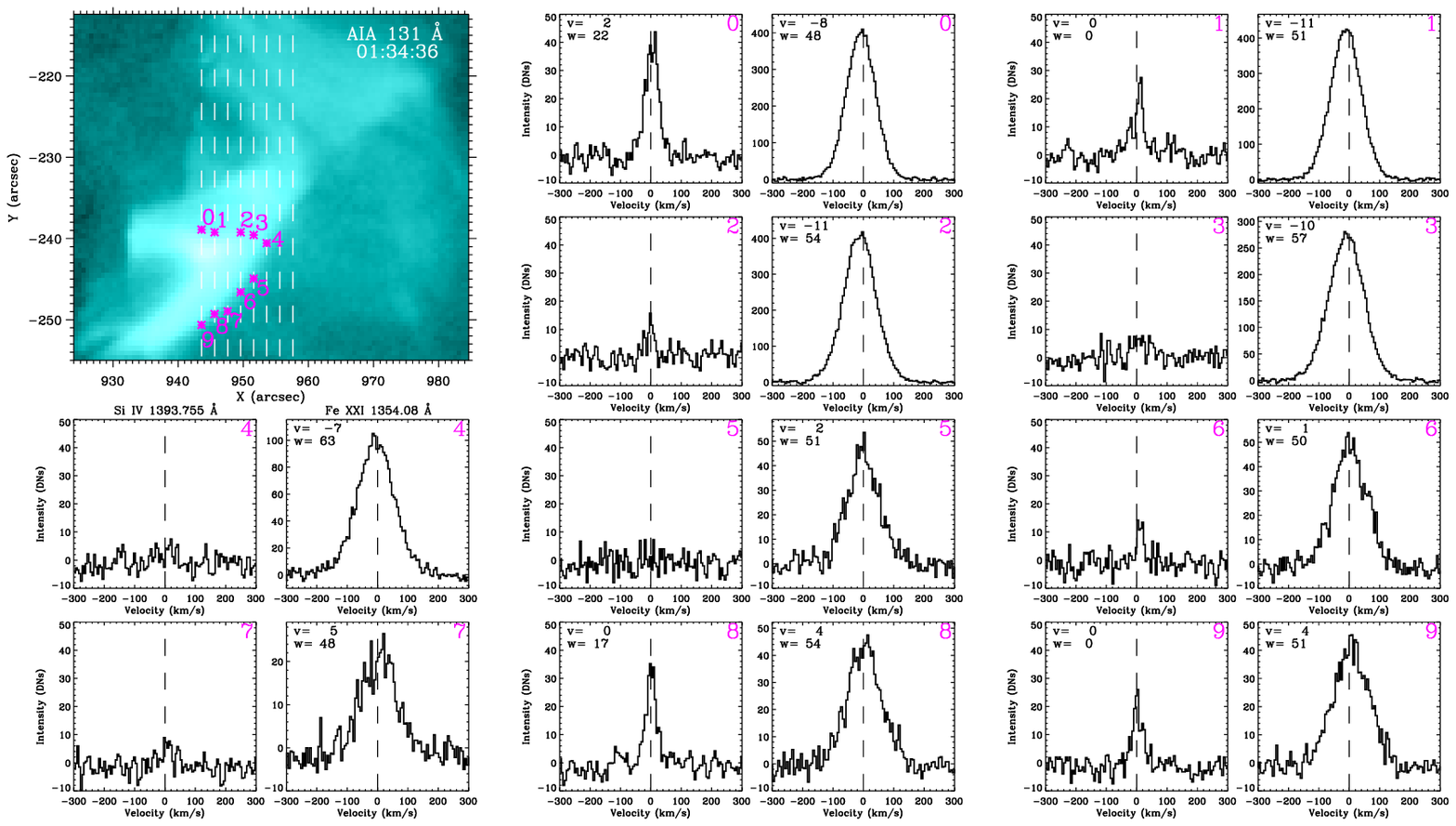}
\caption{
\siiv~1393.76 {\AA} and Fe {\sc xxi} 1354.08 {\AA} line profiles for ten different locations along a loop structure as shown in the AIA 131 {\AA} image at $\sim$01:34 UT on 2014 October 30. These locations correspond to the ones marked in the fifth column of Figure \ref{fig-9} and are labeled as 0--9. The vertical dashed line in the panels of line profiles indicates the reference wavelength of the \siiv\ or Fe {\sc xxi} line. The Doppler velocity and line width (in units of km s$^{-1}$) derived from the moment analysis are presented. Note that the \siiv\ line shows an enhancement at some locations.}
\label{fig-10}
\end{figure*}

Figure \ref{fig-10} plots the line profiles of \fexxi\ 1354.08 {\AA} and \siiv\ 1393.76 {\AA} along the flare loop structure for one of the ten rater scans (marked by the magenta asterisk symbols in Figure \ref{fig-9}) around the peak time of the M-class flare. One can see evident hot \fexxi\ emissions rather than cool \siiv\ emissions, in particular near the loop-top region (locations 3--5). The \fexxi\ line profiles show a good symmetric Gaussian shape with very small blueshifts (see locations 0--4 from the northern halves of the loops) or redshifts (locations 5--9 from the southern halves of the loops) of less than 10 km s$^{-1}$ mostly. These spectral features are quite different from the ones of the newly reconnected loops. It is also seen that the loop-top region (say, location 4) exhibits a slightly larger width of the \fexxi\ line than the loop legs.

\subsection{Downflows from Coronal rain}
\label{thi-re3}

\begin{figure*}[ht!]
\epsscale{1.2}
\plotone{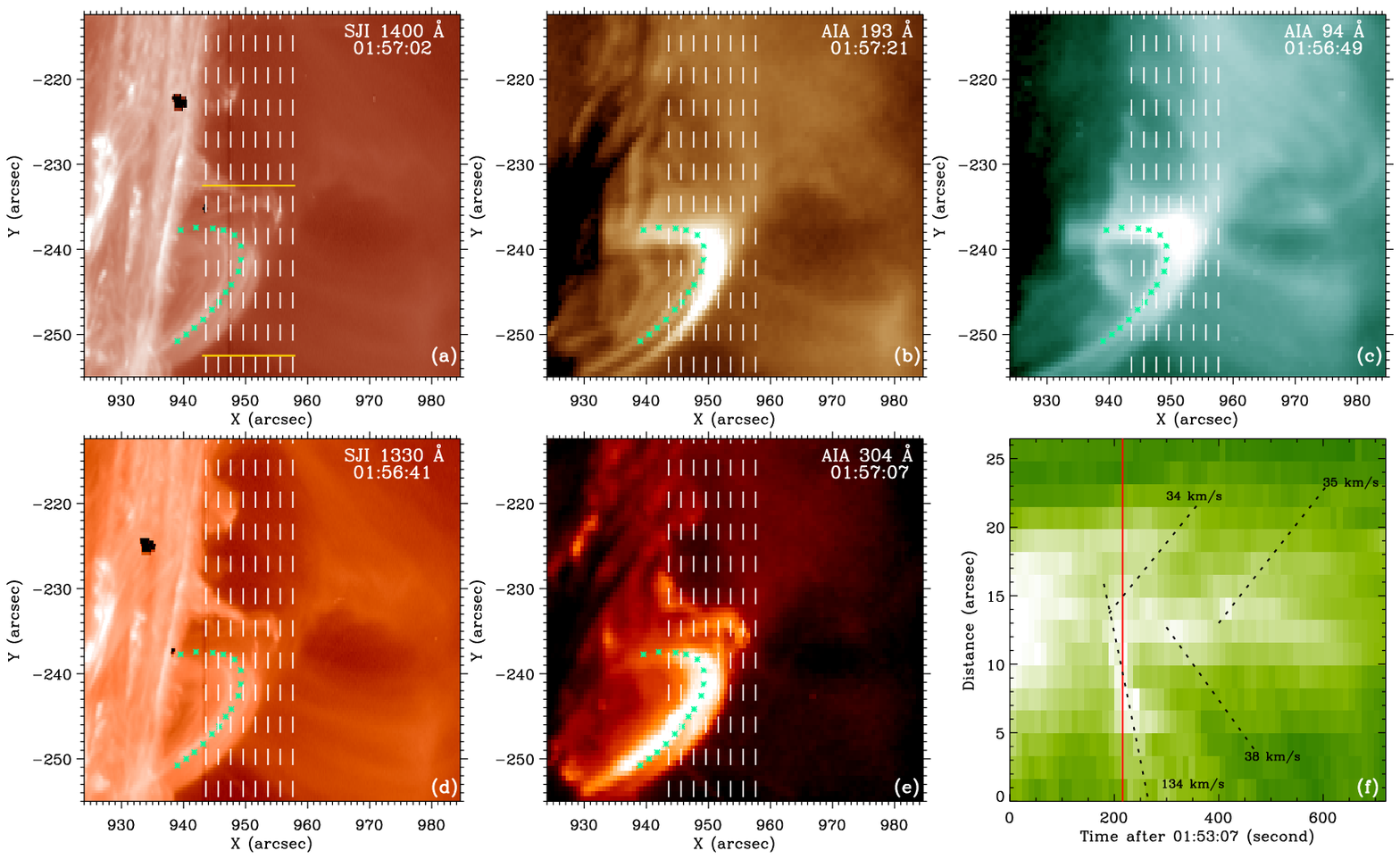}
\caption{
(a)--(e): IRIS SJI 1400 {\AA}, AIA 193 {\AA}, AIA 94 {\AA}, SJI 1330 {\AA}, and AIA 304 {\AA} images taken at $\sim$01:57 UT on 2014 October 30 for the M3.5 flare. The white vertical dashed lines in these panels mark the eight scan steps of the IRIS slit. The two yellow horizontal lines in panel (a) denote the region as shown in Figure \ref{fig-12}. (f) Time-space map at AIA 304 {\AA} for the slice along a loop structure indicated by the green curve in panels (a)--(e). The red vertical line marks the time at $\sim$01:57 UT, corresponding to the one in IRIS and AIA images. Some measured velocities in the plane of the sky are given on the map.}
\label{fig-11}
\end{figure*}

During the decay phase of the M3.5 flare, one can also see some downward flows along the flare loops, which are mainly caused by thermal plasma cooling, called warm coronal rains or coronal condensation. The multi-temperature images in Figures \ref{fig-11}(a)--(e) clearly show both hot and cool flare loops at a late time ($\sim$01:57 UT). The coronal rain draining in the flare loops can be seen from the time-space map of AIA 304 \AA\ along the loop slice (i.e., the green curve in the panel) for the decay period of 01:53--02:15 UT (in Figure \ref{fig-11}(f)). The draining speeds in the plane of the sky are from a few tens of km s$^{^{-1}}$ to more than one hundred km s$^{^{-1}}$, which are comparable with the ones of reconnection downflows that mainly appear in the rise phase of the flare.

\begin{figure*}[ht!]
\epsscale{1.2}
\plotone{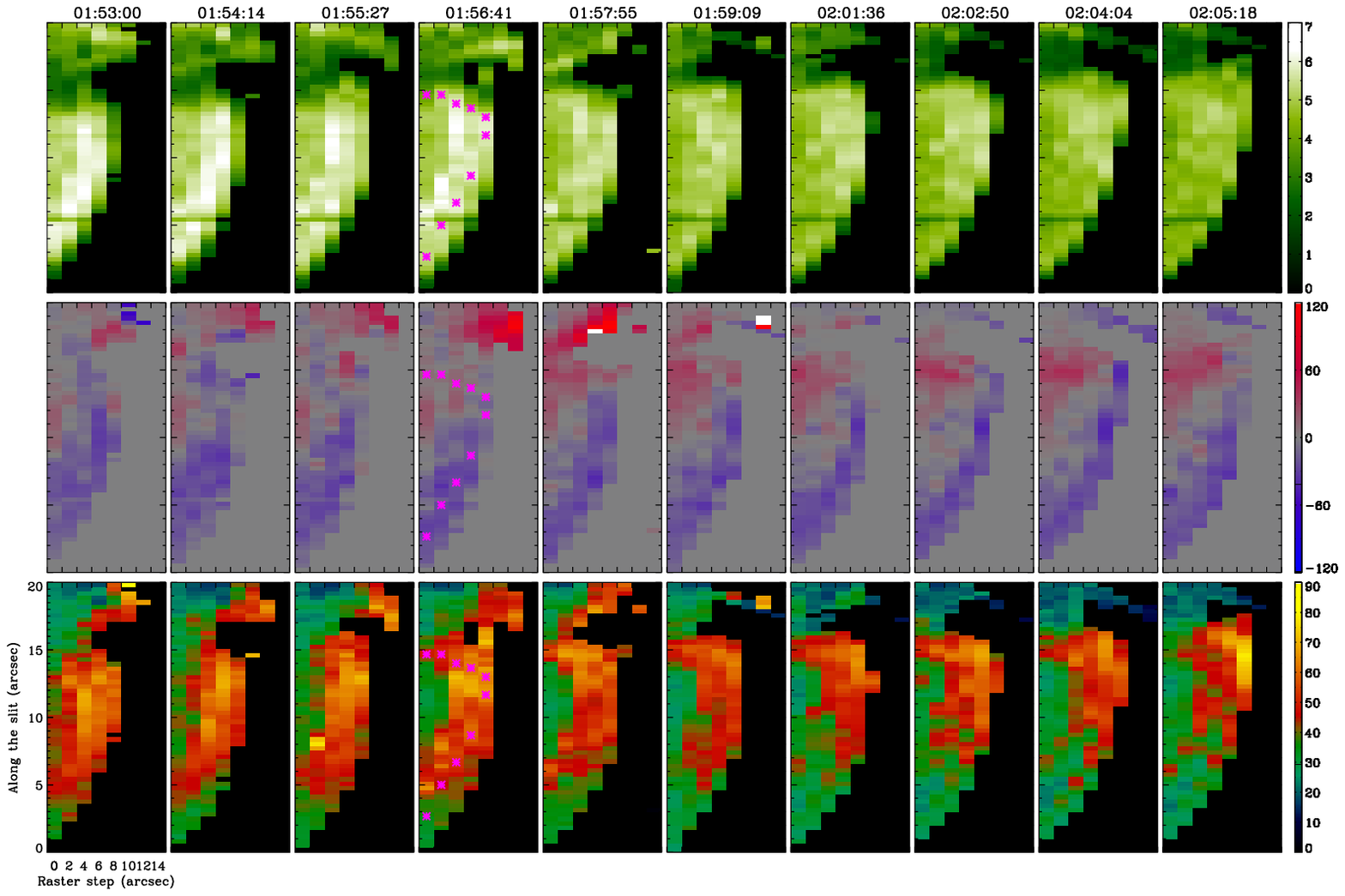}
\caption{
The maps of integrated intensity (top row, in units of log (DNs)), Doppler velocity (middle row, in units of km s$^{-1}$) and line width (bottom row, in units of km s$^{-1}$) of the \siiv~1393.76 {\AA} line obtained from moment analysis from 01:53 UT to 02:05 UT on 2014 October 30 with an interval of 74 s, which corresponds to the time period for the time-space map in Figure \ref{fig-11}(f). The starting time for each raster scan is given at the top of each column. The ten magenta asterisk symbols in the fourth column (corresponding to $\sim$01:57 UT) denote the locations along a loop structure where the \siiv\ line profiles are exhibited in Figure \ref{fig-13}.}
\label{fig-12}
\end{figure*}

From the integrated intensity, Doppler velocity, and line width maps of \siiv\ 1393.76 {\AA} for ten raster scans in Figure \ref{fig-12}, one can also see the warm rain draining process. Similar to the reconnection downflows, the \siiv\ line generally shows redshifts on the northern halves of the loops and blueshifts on the southern halves of the loops, with velocities of a few tens of km s$^{-1}$ along the line of sight. Note that these moment maps look fuzzy, unlike the ones for newly reconnected loops. In addition, the loop-top region has a larger line width than the loop legs.

\begin{figure*}[ht!]
\epsscale{1.2}
\plotone{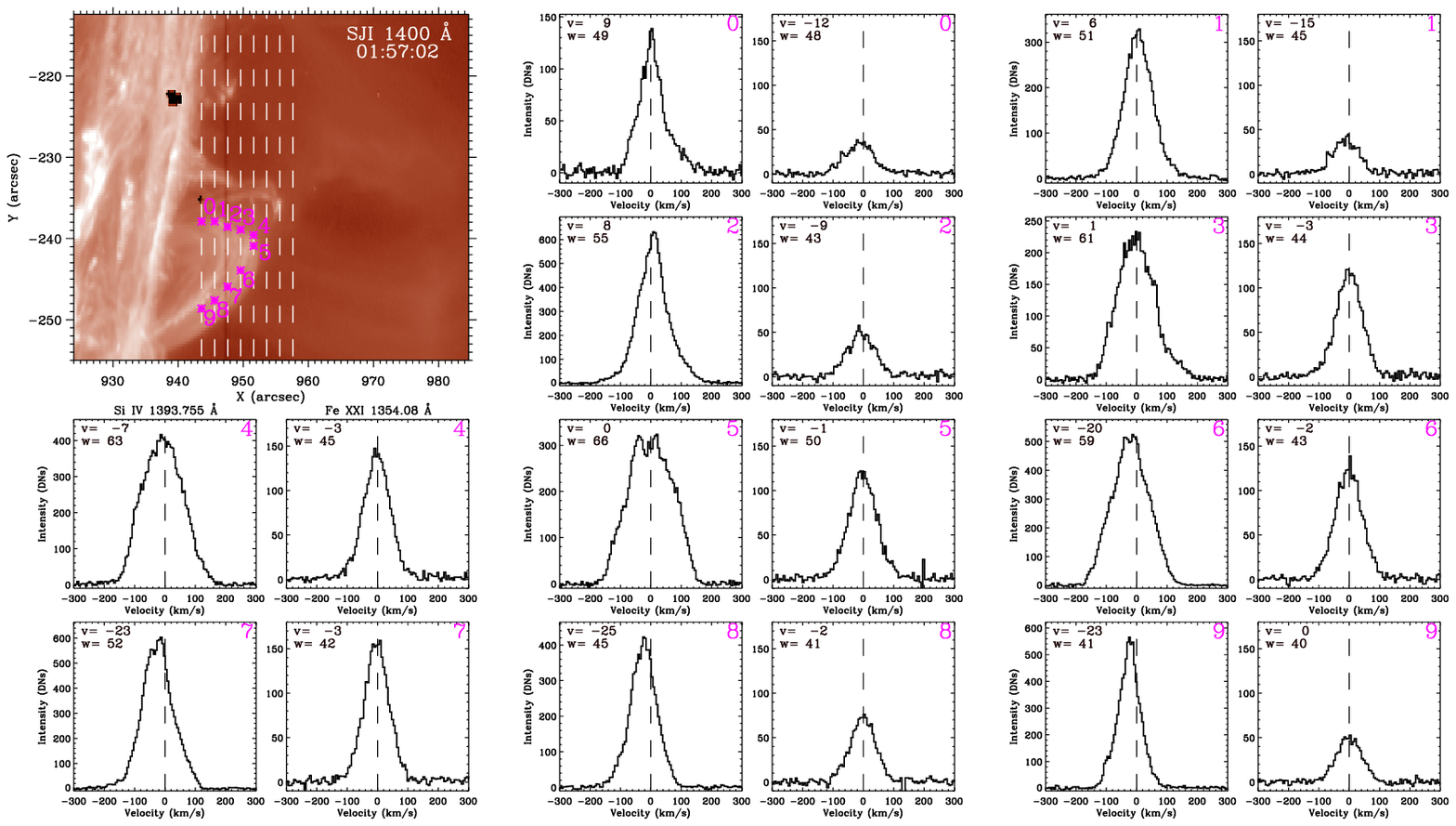}
\caption{
\siiv~1393.76 {\AA} and Fe {\sc xxi} 1354.08 {\AA} line profiles for ten different locations along a loop structure as shown in the IRIS SJI 1400 {\AA} image at $\sim$01:57 UT on 2014 October 30. These locations correspond to the ones marked in the fourth column of Figure \ref{fig-12} and are labeled as 0--9. The vertical dashed line in the panels of line profiles indicates the reference wavelength of the \siiv~or Fe {\sc xxi} line. The Doppler velocity and line width (in units of km s$^{-1}$) derived from the moment analysis are presented.}
\label{fig-13}
\end{figure*}

Figure \ref{fig-13} shows the \siiv\ 1393.76 {\AA} and \fexxi\ 1354.08 {\AA} line profiles along the draining loop structure obtained from one of the ten raster scans. Different from the reconnection and evaporation phenomena, both of the lines exhibit evident emissions for the coronal rains. In addition, the two lines generally show a symmetric shape with some redshifts or blueshifts from several to tens of km s$^{-1}$. Note that near the loop-top region (locations 3--6), the \siiv\ profiles display some small intensity fluctuations around the line center, which may be caused by a superposition of different flow components. This may also be the reason that the line profiles in the loop-top region are wider than the ones near the loop legs.  

\section{Summary and Discussions}
\label{fou-dis}

In this study, we present imaging and spectroscopic observations from SDO/AIA and IRIS for various dynamic processes including magnetic reconnection, chromospheric evaporation, and coronal rain draining in two limb solar flares of C2.7 and M3.5. The IRIS slit crosses over the loop structures at different heights via a raster scan mode, which enables us to study the dynamic processes in flare loops from a side view. Our observational results are summarized as follows.

\begin{enumerate}
\item During the rise phase of the two limb flares, bright and dense loop-top structures with a cusp-like shape can be observed in multi-wavelength images. In particular, in the M-class flare, such a loop-top structure is well co-spatial with the RHESSI 25--50 keV emission source. Intermittent downflows with speeds of tens to hundreds of km s$^{-1}$ in the plane of the sky are detected via the time-space map of AIA 304 \AA. These provide imaging evidence for intermittent magnetic reconnection. Due to the projection effect, the reconnection downflows are manifested as redshifts on the northern halves but blueshifts on the southern halves of the loops in the IRIS \siiv\ 1393.76 {\AA} line, with velocities of about a few tens of km s$^{-1}$ along the line of sight. Along the newly reconnected loops, the \siiv\ profiles significantly vary in shape, being multi-peaked, red-asymmetric or blue-asymmetric. Generally speaking, the loop-top region exhibits multiple flows with different velocities that are reflected from an obviously larger line width of the \siiv\ line when compared with the loop legs. This implies a more turbulent nature of the loop-top region. Note that the hot \fexxi\ 1354.08 {\AA} line shows a trivial emission at the newly reconnected loops.
\item In the impulsive phase (around the peak time) of the M-class flare, prominent hot flare loops show up especially in high-temperature channels. Upward evaporation motions are observed in the time-space map of AIA 131 \AA, whose speeds are close to one hundred km s$^{-1}$ in the plane of the sky. The \fexxi\ 1354.08 {\AA} line is significantly enhanced and the profiles show a good Gaussian shape with small blueshifts or redshifts. The loop-top region has a slightly larger width of the \fexxi\ line than the loop legs. Note that the cool \siiv\ 1393.76 {\AA} line shows a trivial emission at the flare loops with evaporation.
\item During the decay phase of the M-class flare, warm coronal rains can be clearly seen in both high- and low-temperature channels. The downward draining motions are mainly detected on the time-space map of AIA 304 \AA, with velocities of a few to one hundred km s$^{-1}$ in the plane of the sky. The \siiv\ 1393.76 {\AA} line generally shows redshifts on the northern halves and blueshifts on the southern halves of the flare loops due to the projection effect, with velocities of about a few tens of km s$^{-1}$ along the line of sight. Similarly, the loop-top region has a larger width of the \siiv\ line than the loop legs. The profiles of both \siiv\ 1393.76 {\AA} and \fexxi\ 1354.08 {\AA} lines exhibit a relatively symmetric shape but with small intensity fluctuations around the line center.
\end{enumerate}

These observational results reveal some dynamic properties related to magnetic reconnection, chromospheric evaporation, and coronal draining in solar flares. (1) The three processes can last for more than ten minutes and involve multiple loops. In particular, the magnetic reconnection is intermittent in the two limb flares, which might be related to activation of successive loop strands or threads \citep[e.g.,][]{hori98,warr05,bros13} that cannot be spatially resolved. However, we could not rule out another possibility that the intermittent behavior of reconnection downflows originates from one single loop, say, via episodes of energy release. (2) The three phenomena are associated with multi-temperature plasmas that are related to heating and cooling during the evolution of the flare. Specifically speaking, the reconnection downflows can consist of cool plasmas, at least when traveling along the reconnected loops, which is somewhat different from what is always expected. On the other hand, the chromospheric evaporation and coronal draining are mostly related to hot plasmas in flare loops, although the latter may also involve cool plasmas. Note that these results are deducted from limb flares that do not suffer from the background contamination. (3) The three dynamic processes show downward or upward plasma motions along the flare loops, whose real velocities (up to a few hundreds of km s$^{-1}$) can be obtained by combining the time-space maps (velocity component in the plane of the sky) and spectral lines (velocity component along the line of sight). In particular, the loop-top region shows a more turbulent nature, which is mainly caused by superposition of multiple flows. (4) The different dynamic processes make the line profiles greatly vary in shape. The line profiles usually deviate from a single Gaussian shape in magnetic reconnection flows, while they look more Gaussian in chromospheric evaporation and also coronal rains. This is likely due to the fact that the velocities of reconnection flows in the loop-top region do not follow a thermal distribution. (5) The magnetic reconnection, chromospheric evaporation, and coronal draining are prevailing in different flare loops formed in different stages and with different configurations. Nevertheless, in most cases, they cannot be strictly separated due to the projection effect.

Our observations for limb flares validate some of the results from numerical simulations. For example, a dense loop-top structure shows up when magnetic reconnection takes place. This is produced in many flare simulations \citep[e.g.,][]{yoko01,zhao17,ruan20,shen22} and caused by an interaction of the reconnection downflows with magnetic loop arcades. \citet{taka15} made a 2D MHD simulation to study the evolution of flare loops, including many features related to magnetic reconnection. They reported that the shocks above the loop top produced by reconnection downflows are important to determine the thermodynamics of flare loops. \citet{yeji20} performed a 2.5D MHD simulation focusing on the turbulence in the current sheet and loop-top region. They also displayed synthetic observables including AIA fluxes and Fe {\sc xxi} 1354.08 {\AA} line profiles. By comparing our observations with simulations by \cite{yeji20}, we can find some similarities, including the enhanced AIA emission related to the reconnection and flare loops. In fact, recent numerical simulations clearly revealed the turbulent nature of the loop-top region, where the reconnection outflows collide with the flare arcades, producing termination shocks and turbulence there \citep{shen22}. In addition, \citet{taka16} found oscillations above the loop top caused by the reconnection outflow in their simulations. We should also mention a quite different scenario that the multiple flows at the loop-top region, could be caused by the evaporation flows coming from two footpoints of the flare loops, as simulated in \cite{ruan18,ruan19}. The reconnection-induced turbulence and evaporation-induced turbulence may work together, though they are dominant in different stages. This topic deserves further studies.

\acknowledgments
We thank the anonymous referee very much for the detailed suggestions and comments that helped to the manuscript. IRIS is a NASA small explorer mission developed and operated by LMSAL with mission operations executed at NASA Ames Research Center and major contributions to downlink communications funded by the Norwegian Space Center (NSC, Norway) through an ESA PRODEX contract. SDO is a mission of NASA’s Living With a Star Program. The project is supported by NSFC under grants 11733003, 11873095, 11903020, 12127901, and 11961131002, and by National Key R\&D Program of China under grant 2021YFA1600504. Y.L. is also supported by the CAS Pioneer Talents Program for Young Scientists and the CAS Strategic Pioneer Program on Space Science under grants XDA15052200, XDA15320103, and XDA15320301.
         
\bibliographystyle{apj}

\end{document}